\documentclass{ws-procs9x6}

\begin{document}

\title{Testing the HOMOGENEITY OF BRIGHT RADIO SOURCES AT
15~GH\lowercase{z}}

\author{T.~G. Arshakian\footnote{\uppercase{O}n leave from
\uppercase{B}yurakan \uppercase{A}strophysical
\uppercase{O}bservatory, \uppercase{B}yurakan 378433,
\uppercase{A}rmenia} \footnote{\uppercase{TGA} is grateful to the
\uppercase{A}lexander von \uppercase{H}umboldt \uppercase{F}oundation
for the award of a \uppercase{H}umboldt
\uppercase{P}ost-\uppercase{D}octoral \uppercase{F}ellowship.} ,
E. Ros and J.~A. Zensus}

\address{Max-Planck-Institut f\"ur Radioastronomie,\\Auf dem H\"ugel
69, 53121 Bonn, Germany\\ E-mail: tigar@mpifr-bonn.mpg.de}

\author{M.~L. Lister}

\address{Department of Physics, Purdue University,\\525 Northwestern
Ave., W.~Lafayette,\\ IN 47907-2036, USA\\ E-mail:
mlister@physics.purdue.edu}


\maketitle

\abstracts{ A sample of radio-loud active galactic nuclei (AGN) at 2cm
is studied to test the isotropic distribution of radio sources in the
sky and their uniform distribution in space. The sample is complete
flux-density limits of 1.5Jy for positive declinations and 2Jy for
declinations between $0^{\circ}<\delta< -20^{\circ}$. The active
galactic nuclei sample comprises of 133 members. Application of the
two-dimensional Kolmogorov-Smirnov test shows that there is no
significant deviation from the isotropic distribution in the sky,
while the generalised $V/V_{\rm m}$ test shows that the space
distribution of AGN is not uniform at high confidence level
(99.9\%). This is indicative of a strong positive evolution of AGN
with cosmic epoch implying that AGN (or jet activity phenomena) were
more populous at high redshifts. It is shown that the evolution
depends strongly on luminosity: low-luminosity QSOs show a strong
positive evolution, while high-luminosity counterparts do not seem to
show any evolution with cosmic epoch.}

\section{The sample and results of statistical analysis}
We investigate the homogeneity of the flux-density limited sample of
the 2 cm VLBA\footnote{Very Long Baseline Array} survey on the sky and
in the space. The sample is compiled by Lister et al.  (in
preparation; see [1,2]) where the main selection criterion is the
flux-density limit at 15~GHz; all variable sources with galactic
latitude $|b| > 2.5^{\circ}$ and with measured VLBA flux densities
exceeding 1.5 Jy (2 Jy for southern sources) at any epoch since 1994
are included in the sample. The complete sample comprises of 133 radio
sources, all are active galactic nuclei: radio-loud and
core-dominated. Most of them have superluminal radio jets on
parsec-scales. There are 95 quasars, 21 BL Lacs, 9 radio galaxies, and
8 sources with no optical counterparts\footnote{See {\sf
http://www.physics.purdue.edu/astro/MOJAVE/} for more details.}.

We performed a two-dimensional Kolmogorov-Smirnov test to show that
the sample is distributed uniformly on the sky. The generalised
version of $V/V_{\rm m}$ test [3,4] is used to show that the $\langle
V/V_{\rm m} \rangle = 0.589 \pm 0.027$ which is indicative of a strong
positive cosmological evolution of AGN with redshift. The BL Lacs show
a similar trend, but this is not statistically significant because of
the small number of sources. The distribution seems to be uniform for
galaxies, i.e. with no luminosity evolution so far. The plausible
explanation is that all 7 radio galaxies occupy the low redshift
region where the density/luminosity evolution is negligible, but
better statistics would be needed to confirm this result.

To investigate the dependence of radio luminosity on the generalised
$V/V_{\rm m}$ statistic, we divided the sample of quasars in two equal
subsamples above and below the absolute luminosity at 15 GHz, $P_{15}
= 10^{27.9}\,{\rm W\,Hz^{-1}}$. For 46 low-luminosity quasars we find
that $\langle V/V_{\rm m} \rangle = 0.658\pm0.036$ with the confidence
level of 99.96\%, indicative that the distribution of $V/V_{\rm m}$ is
biased towards large values, while for 49 strong sources ($\langle
V/V_{\rm m} \rangle = 0.53\pm0.04$, \, $P=41\%$) no significant
deviation from a uniform distribution is found. The K-S test shows
that these distributions are different at 96\% confidence level.  The
Student t test shows that the mean of $V/V_{\rm m}$ values for low-
and high-luminosity quasars are different at high significance level,
0.011 (98.88\%). The low-/high-luminosity quasars evolve differently
with redshift which is indicative that the cosmic evolution depends
strongly on luminosity.

%
%
%
%


\end{document}